\documentclass[prb,aps,twocolumn,superscriptaddress,10pt,showpacs,longbibliography,floatfix]{revtex4-1}
\usepackage{graphicx}
\usepackage{epstopdf}
\usepackage{color}
\usepackage{amsmath,amssymb}
\usepackage{fixmath}

\usepackage [english]{babel}
\usepackage [autostyle, english = american]{csquotes}
\MakeOuterQuote{"}

\begin{document}

\title{Pressure-Dependent Phase Transitions in Hybrid Improper Ferroelectric Ruddlesden-Popper Oxides}

\author{Gabriel Clarke} 
\affiliation{Department of Chemistry, University of Warwick, Gibbet Hill, Coventry, CV4 7AL,United Kingdom}

\author{Dominik Daisenberger} 
\affiliation{Diamond Light Source Ltd, Harwell Science and Innovation Campus, Didcot OX11 0DE, UK}

\author{X. Luo}
\affiliation{Laboratory for Pohang Emergent Materials, Pohang Accelerator Laboratory and Max Plank POSTECH Center for Complex Phase Materials, Pohang University of Science and Technology, Pohang 790-784, Korea}

\author{S. W. Cheong} 
\affiliation{Laboratory for Pohang Emergent Materials, Pohang Accelerator Laboratory and Max Plank POSTECH Center for Complex Phase Materials, Pohang University of Science and Technology, Pohang 790-784, Korea}
\affiliation{Rutgers Center for Emergent Materials and Department of Physics \& Astronomy, Rutgers University, Piscataway, New Jersey 08854, USA}

\author{Nicholas C. Bristowe} 
\affiliation{Centre for Materials Physics, Durham University, South Road, Durham DH1 3LE, United Kingdom}

\author{Mark S. Senn}
\email{m.senn@warwick.ac.uk}
\affiliation{Department of Chemistry, University of Warwick, Gibbet Hill, Coventry, CV4 7AL,United Kingdom}

\date{\today}

\begin{abstract}

The temperature-dependent phase transitions in Ruddlesden-Popper oxides with perovskite bilayers have been under increased scrutiny in recent years due to the so-called hybrid improper ferroelectricity that some chemical compositions exhibit.  However, little is currently understood about the hydrostatic pressure dependence of these phase transitions. Herein we present the results of a high-pressure powder synchrotron X-ray diffraction experiment and $ab~initio$ calculations on the bilayered Ruddlesden-Popper phases Ca$_{3}$Mn$_{2}$O$_{7}$ and Ca$_{3}$Ti$_{2}$O$_{7}$. In both compounds we observe a first-order phase transition between polar $A2_{1}am$ and non-polar $Acaa$ structures. Interestingly, we show that while the application of pressure ultimately favours a non-polar phase --- as is commonly observed for proper ferroelectrics --- regions of response exist where pressure actually acts to increase the polar mode amplitudes. The reason for this can be untangled by considering the varied response of octahedral tilts and rotations to hydrostatic pressure and their trilinear coupling with the polar instability. 

\end{abstract}

\maketitle

\section{Introduction}

Ferroelectric (FE) materials have been widely studied for their technological applications in sensing, memory devices and beyond.\cite{Scott2007} Understanding structural distortions in functional properties under different conditions is crucial in determining the usability of a given material for a particular device. FE phase transitions are generally well-understood as functions of temperature or chemical doping, with the Pb(Ti/Zr)O$_{3}$\cite{Bhatti2016,Sanjurjo1983,Jaykhedkar2020} and BaTiO$_{3}$\cite{Acosta2017,Decker1989,PRUZAN2002,Mejia-Uriarte2006,Mejia-Uriarte2012,Venkateswaran1998} perovskites (and solid solutions related to them) enjoying a great deal of attention. These materials fall under the class of ‘proper’ FEs, as the induction of a spontaneous, switchable polarisation is a primary order parameter (OP) of the FE phase transition. Upon the application of pressure, proper FEs tend to undergo phase transitions to non-polar structures. For example, BaTiO$_{3}$ transitions from a tetragonal FE phase to a cubic paraelectric (PE) phase at 2~GPa at 300~K.\cite{Bull2021}

However, structural types that exhibit proper FEs comprise only a fraction of all polar materials. Improper FE occurs when the polarisation is a secondary OP of the FE phase transition.\cite{Levanyuk1974,Dvorak1974} In the specific case of hybrid improper ferroelectricity (HIF), the primary order parameter is a pair of non-polar octahedral distortions. These couple to a polar mode in a trilinear coupling mechanism; the polarisation is inhibited if the amplitudes of either of the non-polar distortions or the polar mode are reduced.\cite{Benedek2011} HIF is of particular interest in multiferroic research due to its breaking of the ‘d$^{0}$ rule,’ broadening the library of potential materials to those which include magnetic d$^{n}$ cations.\cite{Bousquet2008,Benedek2012} High-pressure studies of HIF materials are sparser than investigations into proper FE materials, with the n~=~2 Ruddlesden-Popper (RP) Ca$_{3}$Mn$_{2}$O$_{7}$ and Ca$_{3}$Ti$_{2}$O$_{7}$ --- which are otherwise among the most-studied materials at ambient pressure\cite{Benedek2011,Harris2011,Nowadnick2016,Senn2015} --- receiving relatively little attention. Ca$_{3}$Mn$_{2}$O$_{7}$ has been shown to be sensitive to pressure,\cite{Zhu2001,Zhu2002,Ye2018} with several experiments evidencing a phase transition from the polar $A2_{1}am$ phase to a non-polar phase at around 1-1.3~GPa. We are not aware of any  experimental work investigating the high-pressure phase transitions of Ca$_{3}$Ti$_{2}$O$_{7}$, and the mechanism by which pressure couples to polarisation in these materials is not well-understood. 

We have previously shown\cite{Senn2015} that Ca$_3$Mn$_2$O$_7$ exhibits a phase coexistence between the polar $A2_1am$ structure and the non-polar $Acaa$ structure over a wide temperature range. The structural differences between these two phases can be visualised as distortions acting on the aristotype tetragonal $I4/mmm$ phase: the $A2_1am$ phase has undergone both an in-phase rotation of the oxide octahedra and an out-of-phase tilt (with irreps X$_2^+$ and X$_3^-$, respectively) which are coupled to a polar displacement with irrep $\Gamma_5^-$, whereas the $Acaa$ phase only exhibits an out-of-phase rotation of the octahedra (the X$_1^-$ mode). A visualisation of these modes is shown in the SI (Figure S1). By contrast, we have recently shown\cite{Pomiro2020} that Ca$_3$Ti$_2$O$_7$ exhibits a first-order phase transition between $A2_1am$ and $Acaa$. With doping by Sr, this switches to a continuous second-order transition at Ca$_{2.15}$Sr$_{0.85}$Ti$_2$O$_7$.\cite{Pomiro2020} This is made possible by a continuous decrease in the magnitude of the rotation mode associated with the TiO$_6$ octahedra, followed by a rotation in the order parameter direction and decrease in magnitude of the tilt mode. We have recently attempted to probe the behaviour of the distortion modes observed in Ca$_{2.15}$Sr$_{0.85}$Ti$_2$O$_7$ under the effect of an applied electric field, finding an subtle preference for the X$_2^+$ rotation mode to unwind rather than the X$_3^-$ tilt mode.\cite{Clarke2021} These results, coupled with the results of the temperature-based investigations, strongly imply that the X$_2^+$ mode is the softer of the two for this phase.

In this work, we perform synchrotron powder X-ray diffraction experiments on Ca$_{3}$Mn$_{2}$O$_{7}$ and Ca$_{3}$Ti$_{2}$O$_{7}$, using a diamond anvil cell to achieve pressures in excess of 30~GPa. Our results show that, in both phases, at high pressures a centrosymmetric phase with a single condensed out-of-phase rotation (X$_{1}^{-}$) is recovered. However, calculations show that --- contrary to what is expected for proper FEs --- the initial application of pressure actually enhances the polar mode in Ca$_{3}$Ti$_{2}$O$_{7}$.  This result can be understood due to the fact that the octahedral rotations are enhanced at a greater rate than the octahedral tilts are suppressed, leading to an enhancement in the trilinear coupling term with the polarization. 

\section{Experimental Details and Data Analysis}

For high-pressure powder diffraction experiments, the samples used in our previous variable temperature study\cite{Senn2015} were loaded into LeToullec-style membrane diamond anvil cells equipped with Boehler-Almax anvils with 400~$\mu$m culets.  The gasket material was Re, pre-indented to about 50~$\mu$m, with a 200~$\mu$m EDM `drilled' hole to form the sample chamber. The pressure-transmitting medium was Ne, small ruby chips and some Cu were added as pressure gauges. The X-ray beam size was 70~$\mu$m round diameter. Powder X-ray diffraction measurements were performed at Beamline I15 at Diamond Light Source. The beam energy was approximately 29~keV, corresponding to a wavelength of 0.42448(1)~\AA{} refined using a LaB$_{6}$ standard. Diffraction patterns were recorded at intervals as the pressure was increased up to a maximum of 39.86~GPa for Ca$_{3}$Mn$_{2}$O$_{7}$ and 38.48~GPa for Ca$_{3}$Ti$_{2}$O$_{7}$. Pawley refinements were performed against the data using TOPAS Academic v6.\cite{Coelho2018} Compressibility calculations for both phases were performed using the principal axis strain calculator PASCal.\cite{Cliffe2012}

The DFT calculations on Ca$_3$Ti$_2$O$_7$ were performed using the Vienna Ab Initio Simulation Package (VASP), version 5.4.4.\cite{Kresse1993,Kresse1996,Kresse1996b,Kresse1994} We employed the PBEsol exchange correlation potential and projector augmented-wave (PAW) pseudopotentials, as supplied within the VASP package.\cite{Perdew2008,Blochl1994,Kresse1996} A plane wave basis set with a 700 eV energy cutoff and a 6x6x1 Monkhorst-Pack k-point mesh with respect to the parent tetragonal primitive cell (scaled accordingly for other supercells) were found to be appropriate.

\section{Results and Discussion}

\begin{figure}[t]
\includegraphics[width=8.5cm] {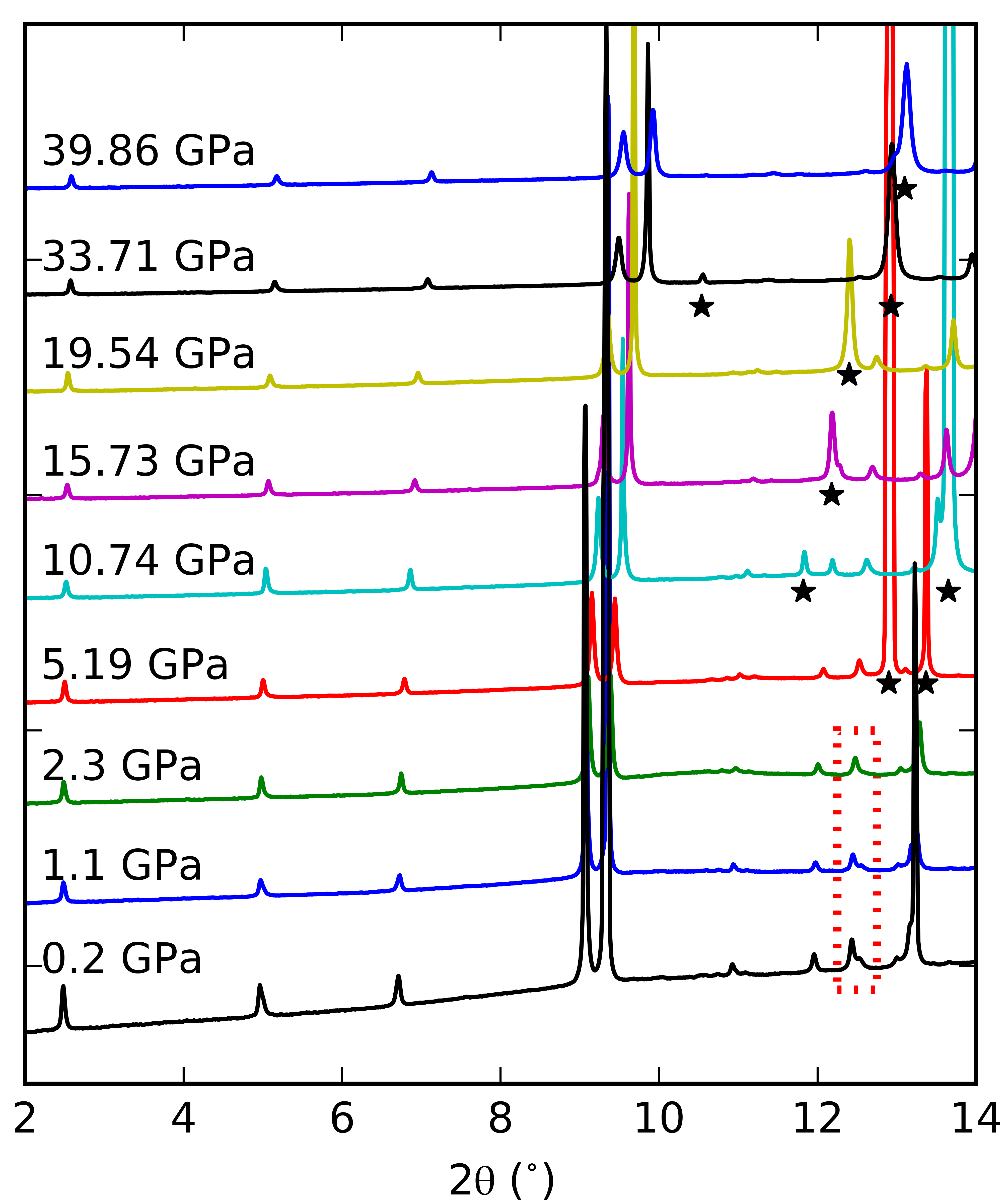}
\caption{\label{Mn_stack} Selected diffraction patterns recorded for Ca$_{3}$Mn$_{2}$O$_{7}$ at a variety of pressures. Diffraction artifacts resulting from the DAC are indicated by stars. The change from observing two (0~0~10) reflections (one each for the $A2_{1}am$ and $Acaa$ phases) to a single reflection for the $Acaa$ phase is indicated by the red dotted rectangle.}
\end{figure}

\begin{figure}[t]
\includegraphics[width=8.5cm] {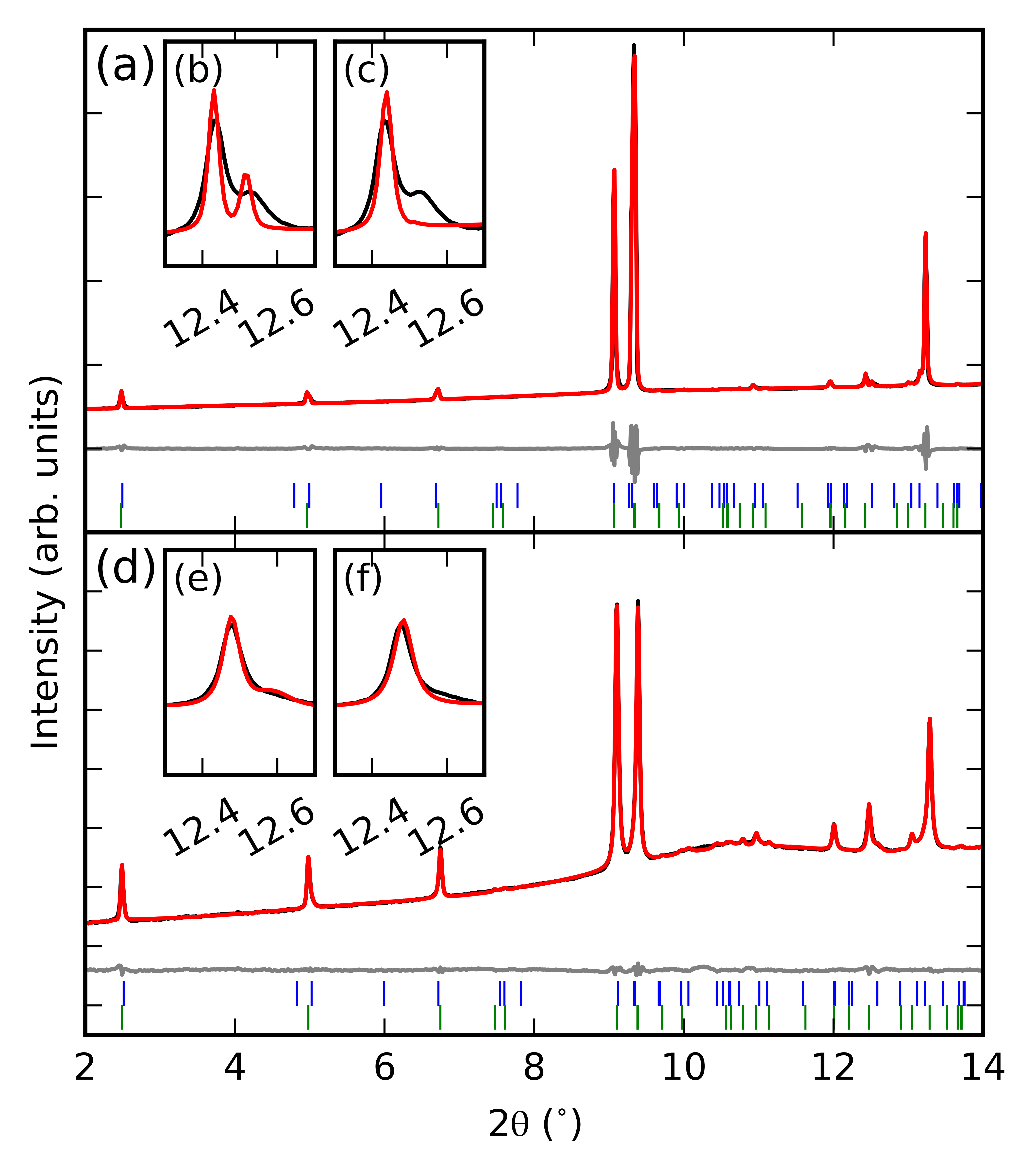}
\caption{\label{Mn_Pawley} Pawley refinement against data gathered for Ca$_{3}$Mn$_{2}$O$_{7}$ at (a) 0.2~GPa and (d) 2.3~GPa showing observed (black), calculated (red) and difference (grey) lines. Upper blue ticks indicate reflections for the $A2_{1}am$ phase, lower green ticks indicate reflections for the $Acaa$ phase. Insets (b) and (e) show the result of a two-phase refinement, showing an improved fit for the (0~0~10) peaks against insets (c) and (f), which show the result of a refinement including only an $Acaa$ phase.}
\end{figure}

We collected high-pressure powder diffraction patterns of Ca$_{3}$Mn$_{2}$O$_{7}$ and Ca$_{3}$Ti$_{2}$O$_{7}$ between 0 and 39.86~GPa. While it is difficult to differentiate the HIF phase $A2_{1}am$ and uniaxial negative thermal expansion (NTE) $Acaa$ phase using Pawley refinements alone, previous work\cite{Senn2016} has shown that the $Acaa$ phase has a consistently longer $c$ axis than the $A2_{1}am$ phase. Assuming that this holds true at the pressures over which we perform our experiment, we may assign the contributions of each phase to the diffraction patterns in Figure \ref{Mn_stack} and \ref{Mn_Pawley}. The occurrence of a phase transition can most easily be appreciated by considering the (0~0~10) reflection associated with the $A2_{1}am$ structure, which disappears around 1~GPa, although the small tail at the high-angle side of the (0~0~10) peak for the $Acaa$ phase indicates that a small amount of highly strained $A2_{1}am$ phase may persist above this pressure.

\begin{figure}[t]
\includegraphics[width=8.5cm] {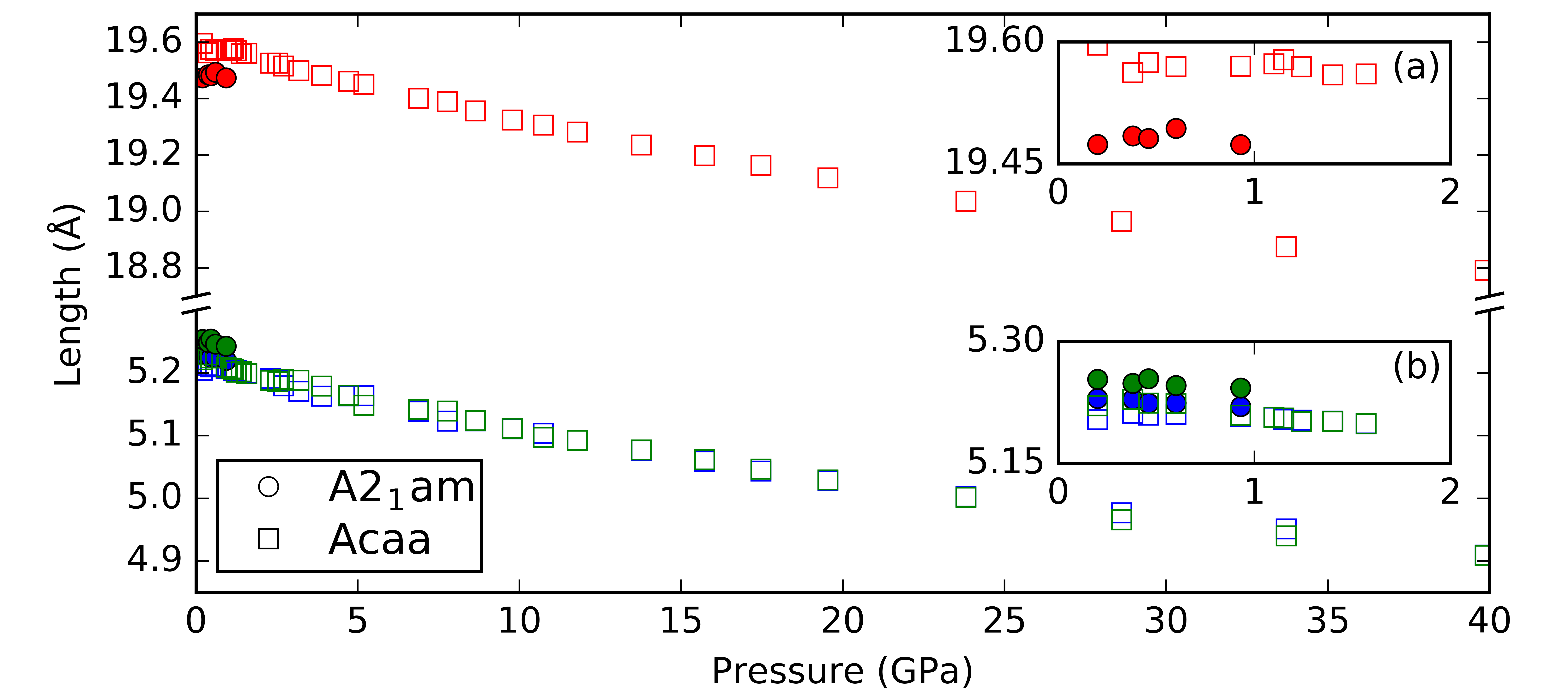}
\caption{\label{Mn_lattice}  Lattice parameters for Ca$_{3}$Mn$_{2}$O$_{7}$. Blue symbols indicate the $a$ parameter, green symbols indicate the $b$ parameter, and red symbols indicate the $c$ parameter. Insets (a) and (b) show the 0-2~GPa regions of each part of the plot, showing detail of the two-phase system.}
\end{figure}

The lattice parameters extracted from Pawley refinements are plotted in Figure \ref{Mn_lattice}, with an inset showing in detail how the two-phase system evolves up to 2~GPa. In contrast to previous work,\cite{Zhu2001} we do not find evidence for a tetragonal-orthorhombic phase transition at 1.3~GPa, nor an orthorhombic-tetragonal phase transition around 9.5~GPa. Instead, the behaviour of the $c$ parameter lends additional support to the dual-phase system at low pressure: single-phase refinements result in a deviation from linearity which is not apparent when the data are refined using the two-phase model described above.

\begin{figure}[t]
\includegraphics[width=8.5cm] {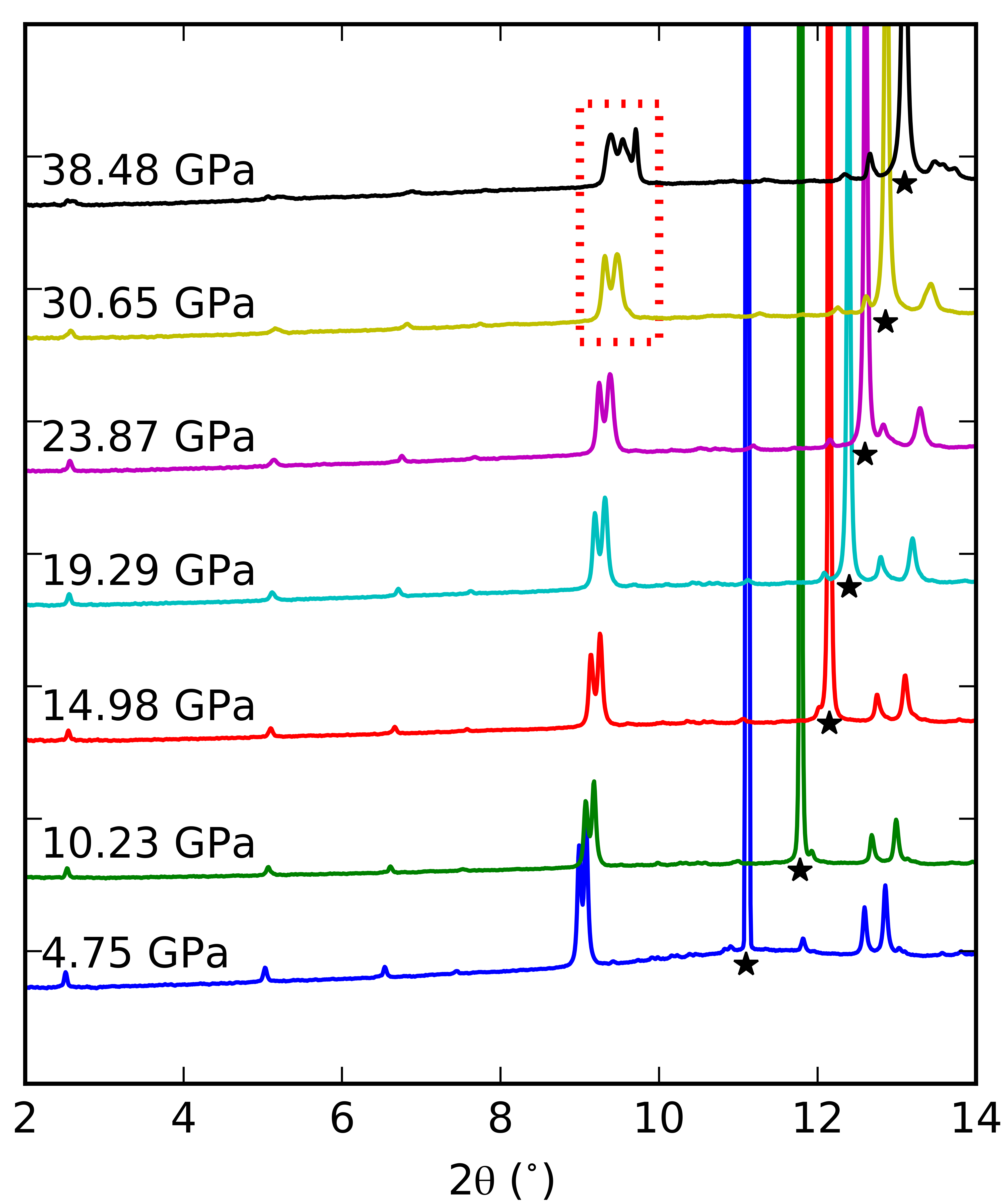}
\caption{\label{Ti_stack} Selected diffraction patterns recorded for Ca$_{3}$Ti$_{2}$O$_{7}$ at a variety of pressures. Diffraction artifacts resulting from the DAC are indicated by stars and the appearance of the (0~2~0) reflection for the $Acaa$ phase is indicated by the red dotted rectangle.}
\end{figure}

\begin{figure}
\includegraphics[width=8.5cm]{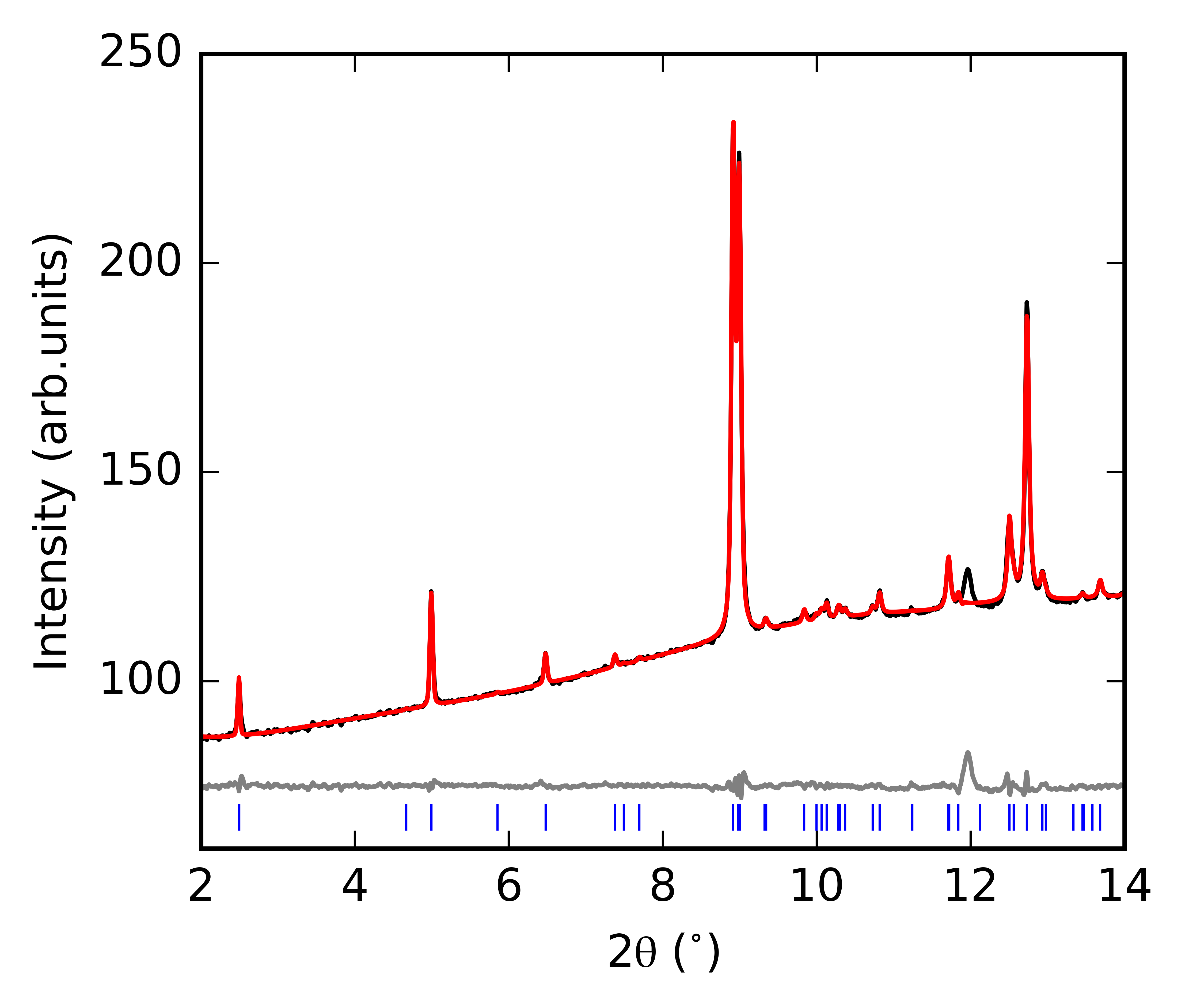}
\caption{\label{Ti_Pawley} Pawley refinement against data gathered for Ca$_{3}$Ti$_{2}$O$_{7}$ at near-ambient pressure, showing observed (black), calculated (red) and difference (grey) lines. Blue ticks indicate reflections for the $A2_{1}am$ phase. The small feature at 2$\theta$~=~12$^{\circ}$ is a result of the diamond anvil cell fulfilling the diffraction condition.}
\end{figure}

\begin{figure}[t]
\includegraphics[width=8.5cm] {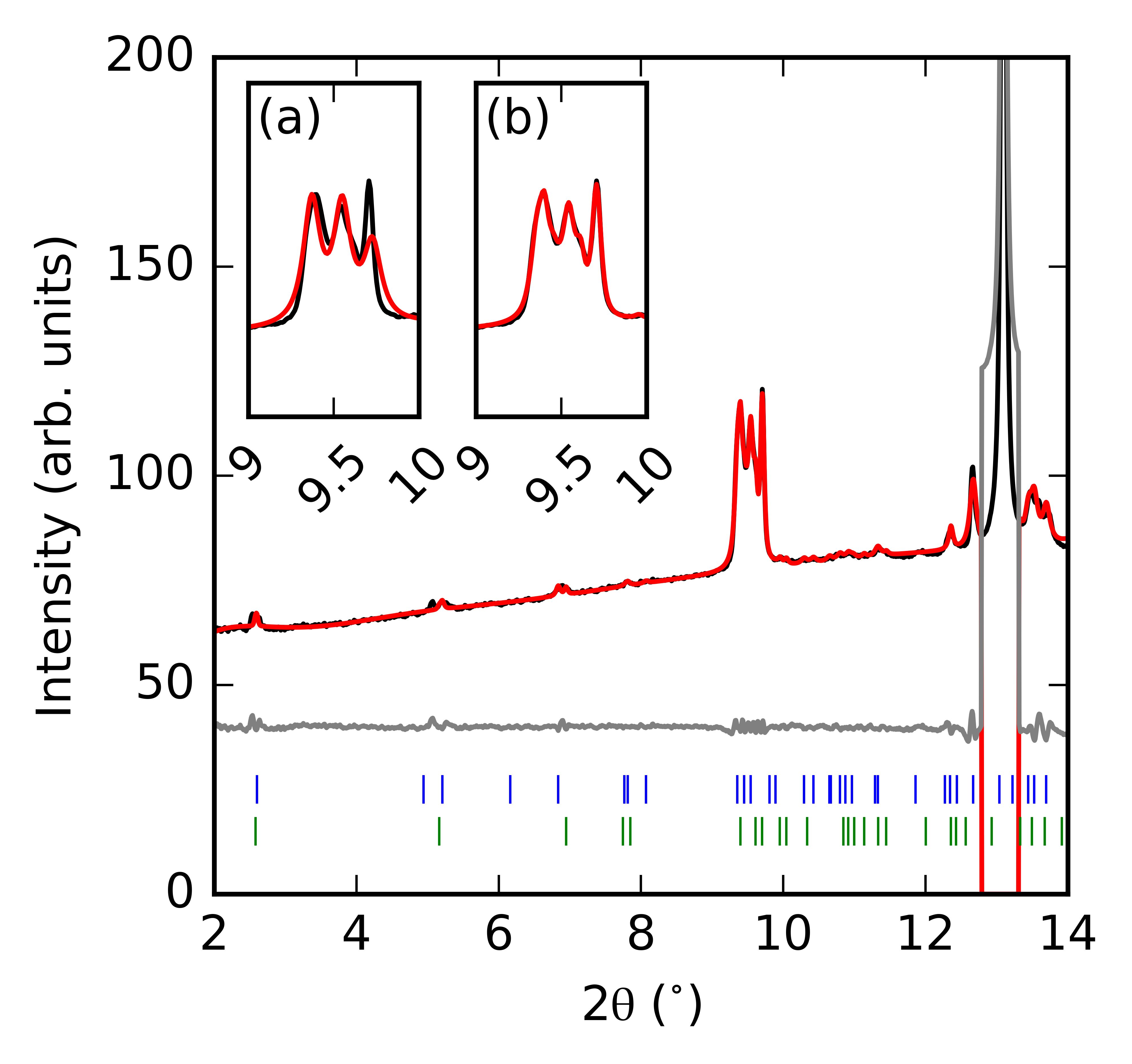}
\caption{\label{Ti_Pawley_HP} Pawley refinement against data gathered for Ca$_{3}$Ti$_{2}$O$_{7}$ at 38.48~GPa, showing observed (black), calculated (red) and difference (grey) lines. Blue ticks indicate reflections for the $A2_{1}am$ phase. Inset (a) shows the result of a single-phase refinement in $A2_{1}am$, inset (b) shows the result of a two-phase refinement including an $Acaa$ phase (green ticks). The feature at 2$\theta$~=~13$^{\circ}$ is a result of the DAC fulfilling the diffraction condition.}
\end{figure}

\begin{figure}[t]
\includegraphics[width=8.5cm] {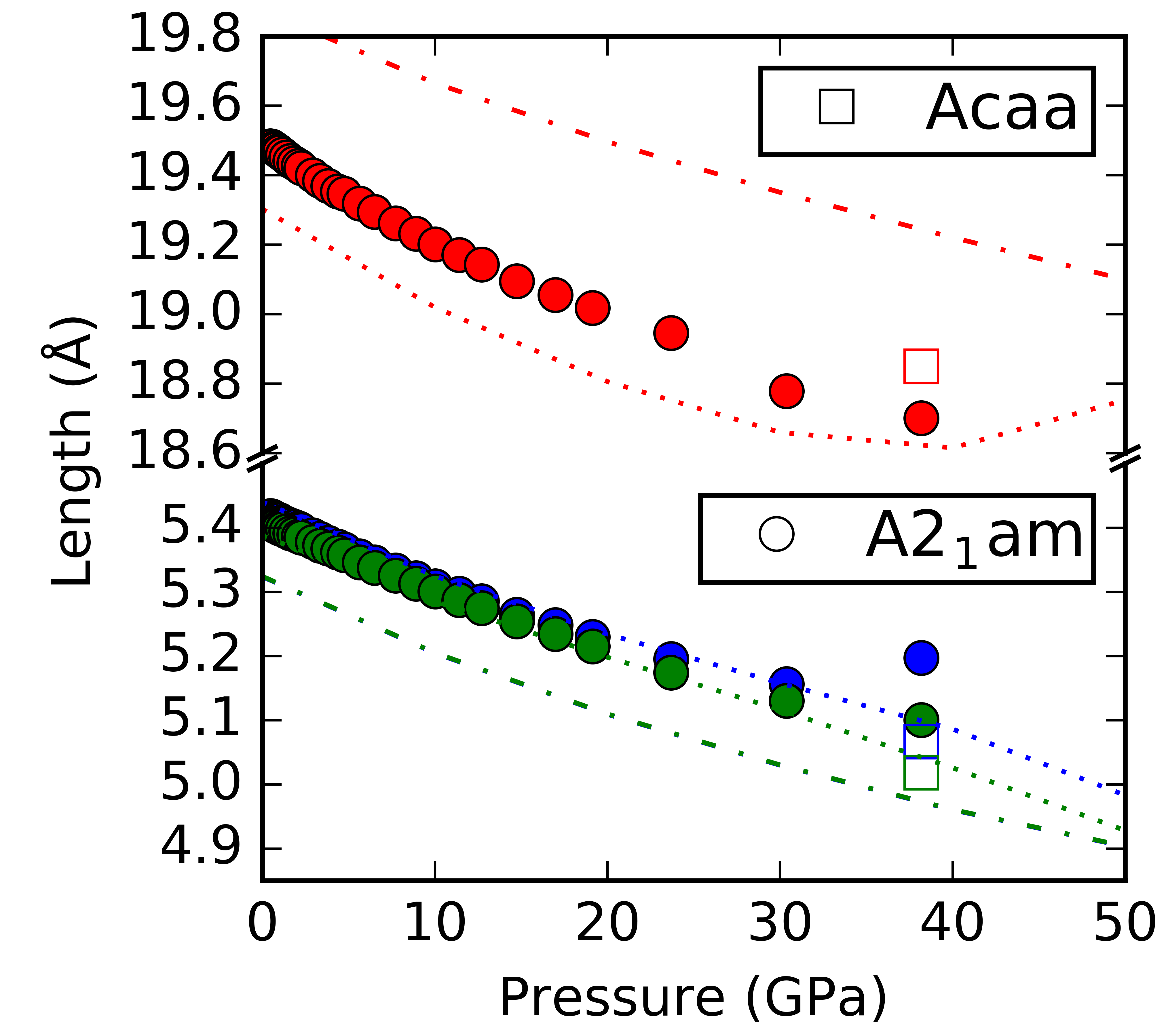}
\caption{\label{Ti_lattice} Lattice parameters for Ca$_{3}$Ti$_{2}$O$_{7}$. Blue symbols indicate the $a$ parameter, green symbols indicate the $b$ parameter, red symbols indicate the $c$ parameter; dotted and dash-dotted lines represent the calculated results for the $A2_{1}am$ and $Acaa$ phases respectively.}
\end{figure}

In contrast to Ca$_{3}$Mn$_{2}$O$_{7}$, Ca$_{3}$Ti$_{2}$O$_{7}$ is well-fit by a Pawley refinement with only a single $A2_{1}am$ phase between ambient pressure and the penultimate pressure point in this experiment (30.65~GPa). Above this pressure, additional reflections appear in the diffraction pattern which we fit by including an $Acaa$ phase. The reflection best-suited to observing this secondary phase is (0~2~0), as highlighted in Figure \ref{Ti_stack} and \ref{Ti_Pawley}. The (0~0~10) reflection is not consistently usable in this case as it becomes contaminated by a reflection from the DAC (Figure \ref{Ti_Pawley_HP}). The experimental and calculated lattice parameters for Ca$_{3}$Ti$_{2}$O$_{7}$ are shown in Figure \ref{Ti_lattice}; they follow a generally linear trend through the aforementioned pressure regime before changing at 38.48~GPa. When extracted from a single-phase $A2_{1}am$ refinement, the $a$ parameter exhibits an upturn at the maximum pressure, while the $b$ and $c$ parameters begin to show signs of plateauing in their otherwise-linear decrease. The DFT-calculated $A2_{1}am$ lattice parameters generally agree well with the experimental results, with the $a$ and $b$ parameters only deviating at the maximum pressure point where there is a phase coexistence. The calculated $c$ parameter is qualitatively similar to its experimental counterpart, though consistently shorter by approximately 0.1~\AA{}, suggesting that ground state DFT calculations overestimate the magnitude of the $X_{3}^{-}$ octahedral tilt observed experimentally at ambient temperature. The calculated $Acaa$ lattice parameters clearly do not agree well with the experiment between 0-30 GPa, indicating that this is not the correct structural model in this pressure regime. The $a$ and $b$ parameters are underestimated and the $c$ parameter is calculated to be longer than the experimental results. However, a qualitative level of agreement across the $A2_{1}am$ to $Acaa$ transition (with $c$ becoming long and $a$ and $b$ shorter) betweeen the experimentally-assigned $Acaa$ phase at 38.48~GPa and the DFT calculations is evident for the final pressure point of our experiment.  The lack of a greater quantitative level of agreement is most likely due to strong strain coupling arising from the intergrowth of  coexisting $A2_{1}am$ and $Acaa$ phases. A plot of the Lorentzian strain extracted from Pawley refinements on Ca$_{3}$Ti$_{2}$O$_{7}$ is shown in Figure S2, and shows who the phase transition acts reduce the microstrain at elevated pressures. Unfortunately, due to experimental limits of the DACs, we were unable to study this sample at higher pressures where a single phase would most likely be realised. The experimental unit cell volumes for Ca$_{3}$Mn$_{2}$O$_{7}$ and Ca$_{3}$Ti$_{2}$O$_{7}$ and calculated unit cell volumes of the latter in $A2_{1}am$, $Acaa$ and $I4/mmm$ are shown in the SI (Figures S3 and S4).

Compressibility parameters and Birch-Murnaghan coefficients were calculated using PASCal \cite{Cliffe2012} using the single-phase data for each composition: the data at pressures greater than 2.3~GPa for Ca$_{3}$Mn$_{2}$O$_{7}$ ($Acaa$ structure) and all data below 30.65~GPa for Ca$_{3}$Ti$_{2}$O$_{7}$ ($A2_{1}am$ structure). These data are tabulated in the SI (Tables S1 and S2). For both compositions, the principal axes are mapped directly on to the crystallographic axes as the tetragonal and orthorhombic structures both have orthogonal lattice vectors by definition. Ca$_{3}$Mn$_{2}$O$_{7}$ is less compressible along all axes (with higher bulk modulus) than Ca$_{3}$Ti$_{2}$O$_{7}$, possibly as a result of the structural differences between the $Acca$ and $A2_1am$ phase. However, any comparisons between the results should be made with caution as the phase transition occurring for Ca$_{3}$Mn$_{2}$O$_{7}$ likely influences the accuracy with which the linear compressibility (extrapolated to zero pressure) are determined.

\begin{figure}[t]
\includegraphics[width=8.9cm] {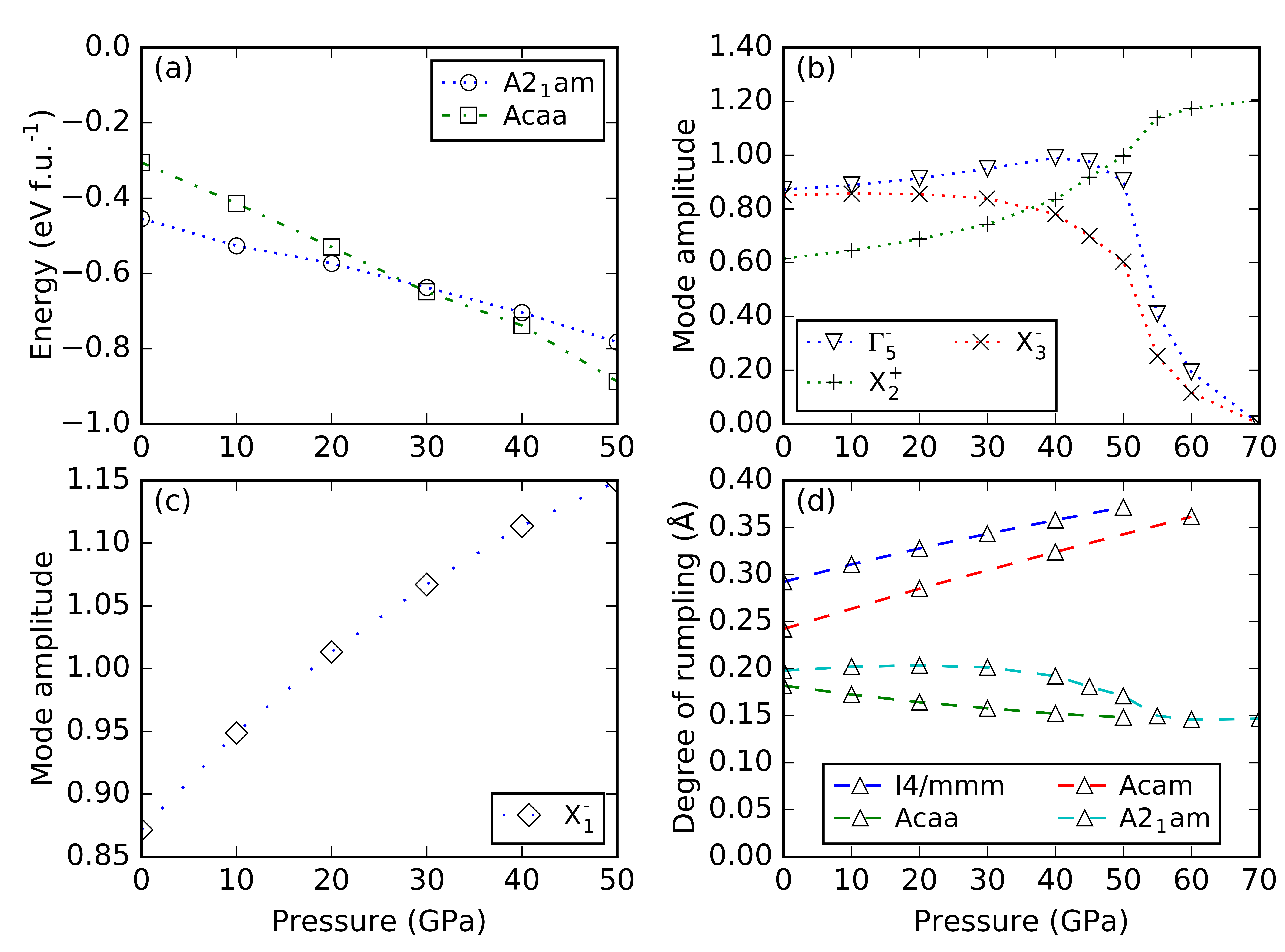}
\caption{\label{DFT} (a): Calculated ground state energies for Ca$_{3}$Ti$_{2}$O$_{7}$ in $A2_{1}am$ and $Acaa$ (relative to $I4/mmm$); (b,c): calculated mode amplitudes ($A_{p}$ values as defined in ISODISTORT) for the $A2_{1}am$ and $Acaa$ structures of Ca$_{3}$Ti$_{2}$O$_{7}$; (d): shows the calculated degree of rumpling for each phase as a function of pressure.}
\end{figure}

Figure \ref{DFT}(a) shows the calculated ground-state energies for the $A2_{1}am$ and $Acaa$ phases of Ca$_{3}$Ti$_{2}$O$_{7}$ relative to the $I4/mmm$ phase up to 50~GPa. DFT finds the polar $A2_{1}am$ phase to be the ground state at 0~GPa, consistent with experimental observations. It can be seen that with increasing pressure, both the polar $A2_{1}am$ and non-polar $Acaa$ phases become increasingly stable with respect to the undistorted $I4/mmm$. However, the $Acaa$ phase is stabilised with pressure at a much greater rate than the $A2_{1}am$ phase, and at approximately 30~GPa it becomes more stable than the polar structure. This is in good agreement with our experimental results wherein we observe the phase transition in the 30.65-38.48~GPa range.

To understand the pressure-induced phase transitions in Ca$_{3}$Ti$_{2}$O$_{7}$ in greater detail, we decompose its DFT-relaxed $A2_{1}am$ and $Acaa$ structures in terms of symmetry-adapted displacements using ISODISTORT\cite{Campbell2006} (Figure \ref{DFT}(b,c)). For the $Acaa$ phases we observe a slow increase in the amplitude of the X$_{1}^{-}$ modes up to 50~GPa, corresponding to the magnitude of the out-of-phase octahedral rotations increasing with pressure. The behaviours of the modes in the $A2_{1}am$ structure were calculated up to 70~GPa and are somewhat different; the X$_{2}^{+}$ in-phase rotation increases in amplitude with increasing pressure up to 40~GPa, while the X$_{3}^{-}$ out-of-phase tilt mode remains constant up to 30~GPa before beginning to decrease in magnitude at 40~GPa. Surprisingly, contrary to what is normally found in proper FEs, the $\Gamma_{5}^{-}$ polar mode that couples to both of these via a trilinear term in the free energy expansion actually increases gradually in this regime. Above 40~GPa, the X$_{3}^{-}$ begins to decrease in amplitude more rapidly, disappearing at 70~GPa, whereas the X$_{2}^{+}$ mode starts to increase more rapidly at this point. We show in Figure S5 that an uncoupled X$_{3}^{-}$ distortion (with the space group $Acam$) increases in amplitude with applied pressure. The polar $\Gamma_{5}^{-}$ mode starts to decreases in magnitude and then disappears finally at 70~GPa, as is to be expected since the trilinear term must necessarily vanish as the tilt magnitude (X$_{3}^{-}$) goes to zero. 

Figure \ref{DFT}(d) shows the degree of rumpling\cite{Zhang2020} of the Ca\nobreakdash-O rock salt-like layer in the $I4/mmm$, $Acaa$ (containing only a rotation distortion with irrep X$_{1}^{-}$), $Acam$ (containing only a tilt distortion with irrep X$_{3}^{-}$) and $A2_{1}am$ phases of Ca$_{3}$Ti$_{2}$O$_{7}$. The term rumpling refers to the loss of coplanarity between the Ca and O atoms in these layers and its magnitude is described by the $\Gamma_{1}^{+}$ distortion mode (that is already active in the $I4/mmm$ parent phase). The rumpling is greatest in the $I4/mmm$ phase and increases with pressure. A similar trend is borne out in the tilt-only $Acam$ phase, while the rumpling is reduced by the inclusion of octahedral rotations (as in the $A2_1am$ structure). These result are in line with a recent study by Zhang \textit{et al.}\cite{Zhang2020} showing that, in n~=~2 RPs, phases with only tilting distortions are observed for larger rumpling amplitudes. Inclusion of  an octahedral rotation (or indeed considering the $Acaa$ phase with only octahedral rotations) leads to a dramatic suppression of the rumpling mode that is exaggerated with increasing pressure. The microscopic origin of this can be understood via the corkscrew mechanism that we have proposed previously\cite{Ablitt2017,Ablitt2018,Ablitt2019} to explain the coupling between in-plane lattice parameter contraction, octahedral rotation, rock-salt layer rumpling and out-of-plane lattice parameter extension, which provides an explanation of the NTE observed in the $Acaa$ phase of these materials.\cite{Senn2015,Senn2016b}

\begin{figure}[t]
\includegraphics[width=8.9cm] {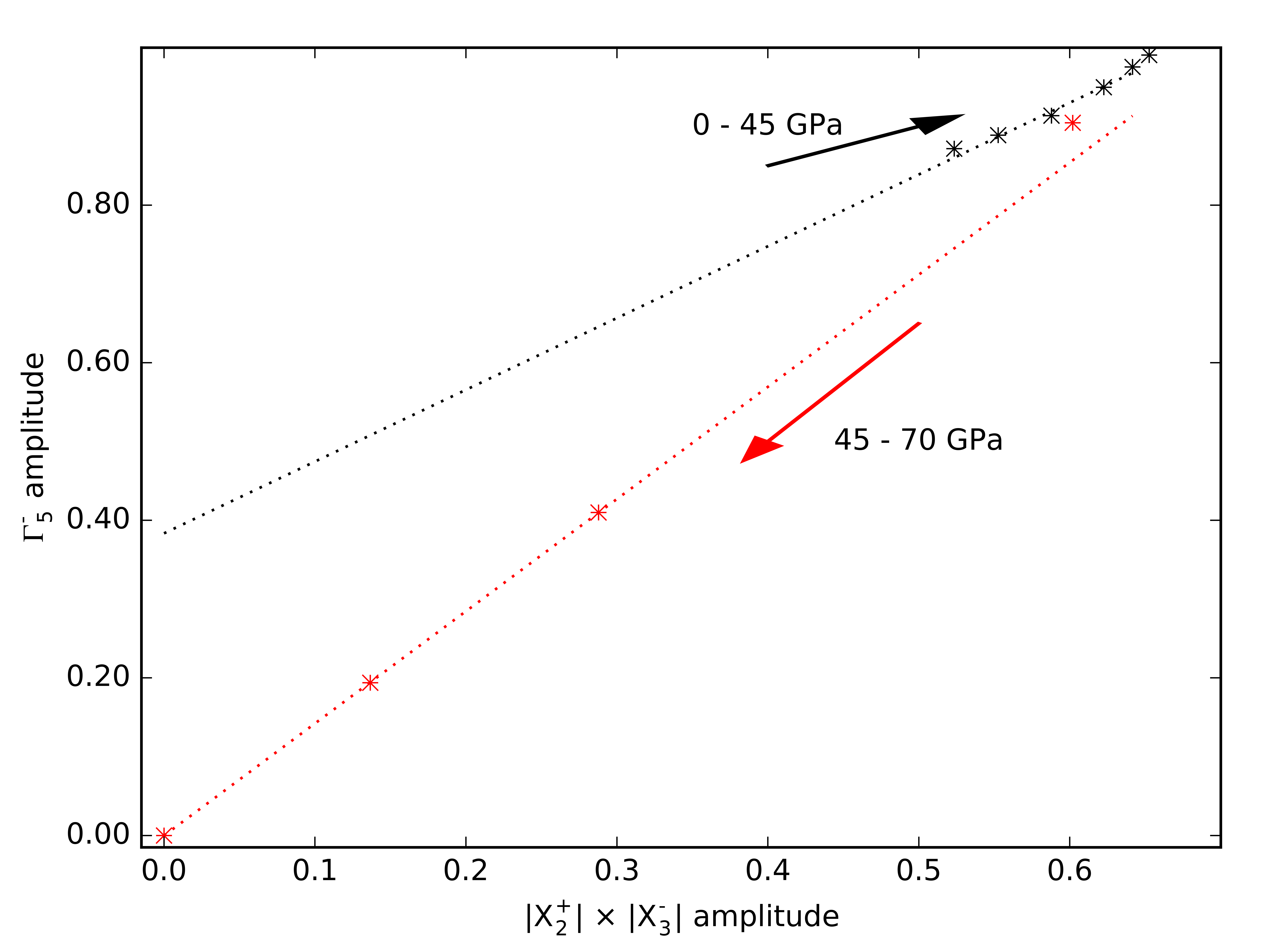}
\caption{\label{tri} Product of the calculated rotation and tilt mode amplitudes plotted against the polar mode amplitude for Ca$_{3}$Ti$_{2}$O$_{7}$, showing deviations of the trilinear coupling mechanism below 30~GPa.}
\end{figure}

To investigate the unexpected trend with pressure that leads to an increase in the polar $\Gamma_{5}^{-}$ mode, we plot its amplitude against that of ($\vert$$X_{2}^{+}$$\vert$~$\times$~$\vert$$X_{3}^{-}$$\vert$) (Figure~\ref{tri}) in order to reveal to what extent this behavior is driven by the trilinear term.\cite{Benedek2011} Individual mode amplitudes are also included in the SI (Figures S7 and S8). This trend indicates that at low pressures the $\Gamma_{5}^{-}$ amplitude is in excess of that expected from the trilinear mechanism, suggesting that there is a small, proper component in addition to that driven by the trilinear coupling.  Only at much higher pressures above 45 GPa is a polarization magnitude consistent with that expected from the trilinear coupling (red line in Figure \ref{tri}) observed. This additional proper component to the polarization is confirmed via DFT calculations on an $F2mm$ phase that shows a $\Gamma_{5}^{-}$ mode which has vanished by a pressure of 40~GPa (See Figure S6).

Comparing our results with the temperature-dependent phase diagrams\cite{Pomiro2020,Senn2015} from the literature we observe that, for both Ca$_{3}$Mn$_{2}$O$_{7}$ and Ca$_{3}$Ti$_{2}$O$_{7}$, elevating the temperature and pressure induces the phase transition from polar $A2_{1}am$ to the centrosymmetric $Acaa$. The fact that both the $A2_{1}am$ and $Acaa$ phases of Ca$_{3}$Ti$_{2}$O$_{7}$ are stabilised simultaneously opposes the normal paradigm observed for proper FE perovskites, where the application of hydrostatic pressure favours the more ‘ordered’ and higher-symmetry ground state and hence behaves analogously to lowering of temperature.\cite{Hayward2005,Janolin2008,Jaouen2007,Jabarov2011,Veithen2002,Toyoura2015}
 
Until recently, relatively little work has been published on the high-pressure behaviours of improper FE materials,\cite{Cross1968,Lucazeau2009,Lucazeau2011,VanAken2004,Jabarov2019,Kozlenko2015} though with the advent of HIF materials in the last decade, a greater interest has been developing. During the preparation of this manuscript, \textit{ab initio} high-pressure calculations by Ramkumar and Nowadnick\cite{Ramkumar2021} up 20 GPa were published on the Ca$_{3}$Ti$_{2}$O$_{7}$, Sr$_{3}$Zr$_{2}$O$_{7}$ and Zr$_{3}$Sn$_{2}$O$_{7}$ phases. All three phases adopt a HIF ground state with the same $A2_{1}am$ symmetry, but with variations in the details of their responses to increasing pressure. As with our findings, in Ca$_{3}$Ti$_{2}$O$_{7}$ a competion between the X$_{3}^{-}$ tilt and the X$_{2}^{+}$ rotation modes was observed that was rationalised as being due to the opposite signs of the coupling of these modes with strain. Although the majority of their study is restricted to the HIF phase only, they do find that the enthalpy of the $Acaa$ phase decreases at a faster rate with respect to the $A2_{1}am$ phase. However, as their calculations are limited to 20 GPa they stop short of the experimental and computational observation of a first-order phase transition that we report here. 

A very recent synchrotron infrared spectroscopy experiment on HIF Sr$_{3}$Sn$_{2}$O$_{7}$ \cite{Smith2021} shows a more complex sequence of pressure-induced phase transitions at 2, 15 and 18~GPa than we observe in Ca$_{3}$Ti$_{2}$O$_{7}$. The first of these is the FE transition, which was assigned as first-order as it resulted from a transition between two phases without a group-subgroup relationship: $A2_{1}am$ and $Pnab$. The $Pnab$ structure still exhibits trilinear coupling, but instead of an in-phase X$_{2}^{+}$ rotation there is an out-of-phase X$_{1}^{-}$ rotation of the oxide octahedra which now couples to antiferrodistortive M$_{5}^{+}$ mode in which the displacements of Sr$^{2+}$ cations cancel one another within a perovskite block. It will be interesting to probe the origin of the transition pathway in this HIF via future high-pressure diffraction studies and DFT calculations.

\section{Conclusion}

The application of hydrostatic pressure to the n~=~2 RP phases Ca$_{3}$Mn$_{2}$O$_{7}$ and Ca$_{3}$Ti$_{2}$O$_{7}$ has been shown to result in a phase transition from the polar $A2_{1}am$ structure to the non-polar $Acaa$ structure via a first order phase transition. These transitions occur at approximately 1~GPa for Ca$_{3}$Mn$_{2}$O$_{7}$ and above 30~GPa for Ca$_{3}$Ti$_{2}$O$_{7}$. \textit{Ab initio} calculations performed for Ca$_{3}$Ti$_{2}$O$_{7}$ indicate that the effect of increasing pressure is not to destabilise the polar structure with respect to the undistorted paraelectric phase, as would be expected for a proper FE material. Instead, the rate at which the $Acaa$ phase is stabilised exceeds the rate for the $A2_{1}am$ phase, leading to a cross-over in the ground state energies at a pressure that is consistent with the experimentally-observed transitions. Crucially, in the $A2_{1}am$ phase of Ca$_{3}$Ti$_{2}$O$_{7}$ we predict a regime where pressure actually acts to increase the overall magnitude of the polar mode. Our results highlight that the response of improper FEs to pressure can be quite different to the conventionally-observed suppression of polarization observed in proper FEs. Consequently, unexplored possibilities exist for using hydrostatic pressure to control the polarization states and switching pathways in improper FEs.

\section{Supporting Information}

Diffraction data and DFT-optimised structures are available as supporting information and may be accessed at https://doi.org/10.6084/m9.figshare.17136548

\begin{acknowledgments}
G.C. was supported by an EPSRC studentship (2020431).  M.S.S. acknowledges the Royal Society for a fellowship (UF160265) and the EPSRC for funding (EP/S027106/1)). N.C.B acknowledges computational resources from the Hamilton HPC Service of Durham University and the UK Materials and Molecular Modelling Hub (partially funded by the EPSRC project EP/P020194/1). The sample was characterised using the “Oxford/Warwick Solid State Chemistry BAG to probe composition-structure-property relationships in solids” (CY25166). The high pressure diffraction study was performed at I15, Diamond Light Source under proposal number EE12312-1; all DACs and sample loadings were provided by DLS I15. MSS would like to acknowledge Dr Claire Murray and Ms Sarah Craddock for assistance in collecting the high pressure data. The work in Korea was supported by the National Research Foundation of Korea (NRF) funded by the Ministry of Science and ICT(No. 2016K1A4A4A01922028), and the work at Rutgers University was supported by the DOE under Grant No. DOE: DE-FG02-07ER46382.
\end{acknowledgments}

\bibliographystyle{apsrev4-1}
\bibliography{references}
\end{document}